\documentclass[prd,twocolumn,showpacs,preprintnumbers,amsmath,amssymb]{revtex4}      
\usepackage{epsfig}%
\usepackage{longtable}
\begin{document}
\setlength{\topmargin}{-0.25in}
\preprint{NSF-KITP-05-23}

\title{About a possible 3rd order phase transition at $T=0$ in 4D gluodynamics}

\author{L. Li}
\affiliation{Department of Physics and Astronomy\\ The University of Iowa\\
Iowa City, IA 52242 \\ USA
}

\author{Y. Meurice}
\altaffiliation{Also at the Obermann Center for Advanced Study}
\email[]{yannick-meurice@uiowa.edu}
%
\affiliation{Kavli Institute for Theoretical Physics, Santa Barbara\\
Santa Barbara, CA 93106 USA \\and \\
Department of Physics and Astronomy\\ The University of Iowa\\
Iowa City, IA 52242 \\ USA}

\date{\today}

\begin{abstract}

We revisit the question of the convergence of lattice perturbation theory for a pure $SU(3)$ lattice gauge theory in 4 dimensions. Using a series for the average plaquette 
up to order 10 in the weak coupling parameter $\beta^{-1}$, we show that 
the analysis of the extrapolated ratio and the extrapolated slope suggests the possibility of a non-analytical power behavior of the form $(1/\beta -1/5.7(1))^{1.0(1)}$, 
in agreement with another analysis based on the same asumption.
This would imply that the third derivative of the free energy density diverges near $\beta =5.7$. 
We show that the peak in the third derivative of the free energy present 
on $4^4$ lattices disappears if the size of the lattice is increased isotropically 
up to a $10^4$ lattice. 
On the other hand, on $4\times L^3$ lattices, a jump in the third 
derivative persists when $L$ increases, and follows closely the known values of $\beta_c$ for the first order finite temperature transition. We show that the apparent contradiction 
at zero temperature can be resolved by moving the singularity in the complex $1/\beta$ plane. If the imaginary part of the location of the singularity $\Gamma$ is within the range $0.001< \Gamma < 0.01$, it is possible to limit the second derivative of $P$ within an acceptable range without affecting drastically the behavior of the perturbative coefficients. We discuss the possibility of checking the existence of these complex singularities by using the strong coupling expansion or calculating the zeroes of the partition function.
\end{abstract}

\pacs{11.15.-q, 11.15.Ha, 11.15.Me, 12.38.Cy}

\maketitle

\section{Introduction}
It is widely accepted that the continuum limit of asymptotically free lattice gauge theories at zero temperature is obtained in the limit of arbitrarily small coupling. 
In practice, however, the relevant information about the continuum is obtained from a 
crossover region where both weak and strong coupling expansions break down, but where 
one can observe the onset of asymptotic scaling. In this region, observables such as the 
average of the 
elementary plaquette are sensitive to the contributions of large field configurations.
Such contributions can be modified by adding an adjoint term \cite{bhanot81,heller95,deforcrand02,hasenbusch04}, a monopole chemical potential \cite{mack78,brower81} or by removing 
configurations with an action larger than some given value \cite{lattice04}. 
These studies illustrate the fact that non-universal features or lattice artifacts seem 
generically present in the crossover region.

On the other hand, one may hope that the universal features of the continuum limit may be obtained directly from a weak coupling expansion. However, 
convergence issues need to be considered.
The discontinuity in the plaquette average \cite{gluodyn04} in the limits $g^2\rightarrow 0^{\pm}$ precludes the existence of a regular perturbative series, and the 
decompactification of the gauge variables in lattice perturbation \cite{heller84}  
should lead to asymptotic series \cite{plaquette}. Despite these considerations, an 
analysis of the first 10 coefficients for the average plaquette $P$ for the standard Wilson action in the fundamental representation \cite{heller84,alles93,direnzo95,direnzo2000} suggests \cite{rakow2002} a finite 
radius of convergence and a non-analytic behavior of the form
\begin{equation}
P\simeq A_0(1/\beta _c -1/\beta )^{1-\alpha } \ .
\label{eq:convpar}  \end{equation}
In the present context, $\alpha \simeq 0$. If $\alpha>0$, the first derivative of $P$ diverges at $\beta_c$ and we say that the transition is second order since $P$ is obtained by taking the derivative of the free energy with respect to $\beta$. If $-1<\alpha<0$, the first derivative is bounded but the second derivative of P diverges and we call the transition third order.
This type of singularity is not expected for the model considered here (no adjoint term in the action). The divergence at $\beta_c$ requires long range correlations and consequently a massless state. No such a state has been 
observed in glueball spectrum studies \cite{berg82,michael88,bali93}. Consequently, 
a natural strategy would be to try to falsify Eq. (\ref{eq:convpar}).

In this article, we test the validity of Eq. (\ref{eq:convpar}) by directly calculating the first and second derivative of $P$ near the hypothetical $\beta_c$. 
We  consider the minimal, unimproved, lattice gauge model originally 
proposed by K. Wilson \cite{wilson74c}. 
With standard notations, the lattice functional integral or partition function is 
\begin{equation}
\label{eq:z}
Z=\prod_{l}\int dU_l {\rm e}^{-\beta \sum_{p}(1-(1/N)Re Tr(U_p))} \,
\end{equation}
with $\beta=2N/g^2$.
Our study focuses on $P$ and its derivatives which are defined more precisely in Sec. \ref{sec:direct}. All numerical calculations are done with the fundamental representation of 
$SU(3)$.

In Sec. \ref{sec:seran}, we discuss various ways to estimate the unknown parameters in 
Eq. (\ref{eq:convpar}) using the perturbative sereis. All the methods 
give results which are reasonably consistent and show the robustness of the analysis of 
Ref. \cite{rakow2002}. 
In all cases, $\alpha$ is very close to 0. If we use the scaling relation $\alpha=2-D\nu$ with $D=4$, $\alpha=0$ goes with the mean field result $\nu = 1/2$. Note that $\alpha =0$ is borderline between second and third order transition. In the mean field theory of spin models, the specific heat has a discontinuity. It is common to associate a discontinuity or a logarithmic divergence of the specific heat with a second order phase transition.  
At the other end, a 
discontinuity in the 
derivative of the specific heat is often called a third order phase transition. It has been observed in the context of the large-$N$ limit of gauge theories in two dimensions \cite{gross80}, random surface models \cite{catterall88} and spin glasses \cite{crisanti}. 
In $3D$ $O(N)$ models, $\alpha \simeq 0.11$ for $N=1$ (second order) and -0.12 for $N=3$ (third order). 
In Sec. \ref{sec:remarks}, we discuss the possibilty of having asymptotic series behaving temporarly as series with a finite radius of convergence and we discuss the effect of the tadpole improvement 
\cite{lepage92}.
In Sec. \ref{sec:direct}, we calculate the first two derivatives of $P$ on $L^4$ lattices.
and also on $4L^3$ lattices together with the average of the Polyakov loop. 
In Sec. \ref{sec:resolution}, we show that the absence of peak increasing with the volume 
for the second derivative of $P$ can be accomodated  by moving the singularity in the complex $1/\beta$ plane. If the imaginary part of the location of the singularity 
is kept reasonably small but not too small, it is possible to limit the second derivative of $P$ within an acceptable range without affecting drastically the behavior of the perturbative coefficients. In Sec. \ref{sec:checks},  we discuss the possibility of checking the existence of these complex singularities by using the strong coupling expansion \cite{balian74err,falcioni80,falcioni81} or calculating the zeroes of the partition function 
\cite{falcioni82,alves91,janke05}.

\section{Series Analysis}
\label{sec:seran}
\begin{figure}
\includegraphics[width=2.9in,angle=0]{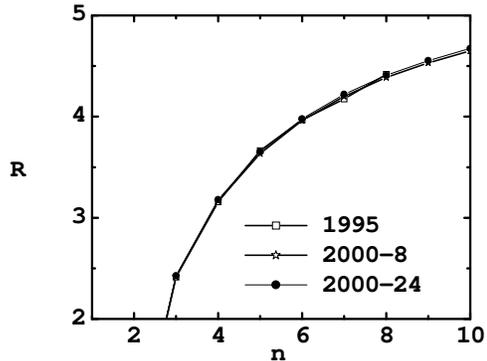}
\caption{$r_m$ versus $m$ for the three data set given in the text.}
\label{fig:rat}
\end{figure}

In this section we analyze the weak coupling series
\begin{equation}
P(\beta)\sim \sum_{m=1}^{10} b_m \beta^{-m} .
\end{equation}
We define $r_m=b_m/b_{m-1}$, the ratio of two
successive coefficients. The $r_m$ are displayed in Fig. \ref{fig:rat}.
In this Fig. and the following, we compare three data sets. The first set \cite{direnzo95} is the 1995,  
order 8, series (empty circles) and the two other sets are the more recent\cite{direnzo2000} order 
10 series calculated on a $8^4$ (stars) and $24^4$ (filled circles) lattices. It is difficult to distinguish the ratios corresponding to the three data sets on Fig. \ref{fig:rat}. The first three coefficients are in good agreement with analytical results 
\cite{heller84,alles93}. 
The apparent convergence of the ratios suggests \cite{rakow2002} that we consider a leading non-analytical behavior the form given in Eq. (\ref{eq:convpar}).
This will be our working hypothese for the rest of the section.

One can estimate the unknown parameters \cite{gaunt} with a fit
$r_m =\beta_c(1+(\alpha -2 )/m)$
For instance, the range 3 to 10 for the ratios gives $\beta_c\simeq 5.66$ and $\alpha\simeq 0.24$ while 
the range 4 to 10 gives $\beta_c\simeq 5.69$ and $\alpha\simeq 0.18$. The $1/m^2$ corrections were reduced with a modified form $r_m =\beta_c(1+(\alpha -2 )/(m+s))$ in Ref. \cite{rakow2002} with a value $s=0.44$. 
Using the range 4 to 10 again, one 
obtains $\beta_c\simeq 5.78$ and $\alpha\simeq-0.01$. The estimates quoted in Ref. 
\cite{rakow2002} are $\beta_c\simeq6\times 0.961\simeq 5.77$ and $\alpha\simeq 0.01$.

\begin{figure}
\includegraphics[width=2.9in,angle=0]{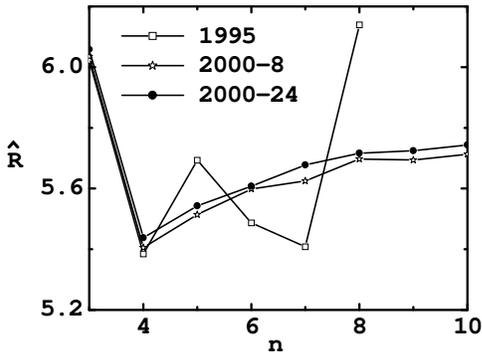}
\caption{$\widehat{R}_m$ versus $m$ for the three data sets described in the text.}
\label{fig:rhat}
\end{figure}

Another way to reduce the uncertainties of order $1/m$ in the estimation of $\beta_c$ 
is to use \cite{nickel80} the extrapolated ratio ($\widehat{R}_m$) defined as
\begin{equation}
\widehat{R}_m = mr_m -(m-1)r_{m-1}\ .
\label{eq:rhat}\end{equation}
The values of $\widehat{R}_m$ are shown in Fig. \ref{fig:rhat}. As the calculation involves differences of successive ratios mutiplied by the order, the result is sensitive to statistical errors. Fig. \ref{fig:rhat} illustrates the fact that the 2000 series has much smaller statistical errors than the 1995 series. Fig. \ref{fig:rhat} also indicates that as the volume increases from $8^4$ 
to $24^4$, $\widehat{R}_m$ increases by approximately 0.03 which is slightly larger than the changes in $\widehat{R}_m$ with $m\leq8$. 
Our best estimate of $\beta_c$ is $\widehat{R}_{10}=5.74$ for the $24^4$ lattice. 
Given the other estimates and the size of the volume effects, it is reasonable to conclude that $\beta_c=5.7(1)$.

\begin{figure}
\includegraphics[width=2.8in,angle=0]{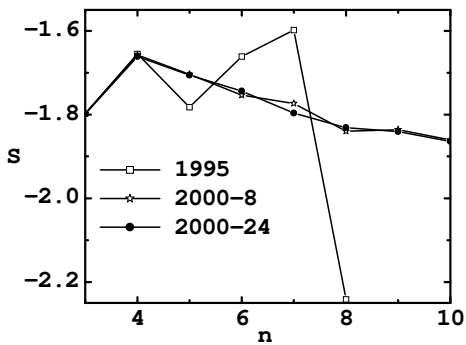}
\caption{The normalized slope $S_n$.}
\label{fig:noslo}
\end{figure}

\begin{figure}
\includegraphics[width=2.8in,angle=0]{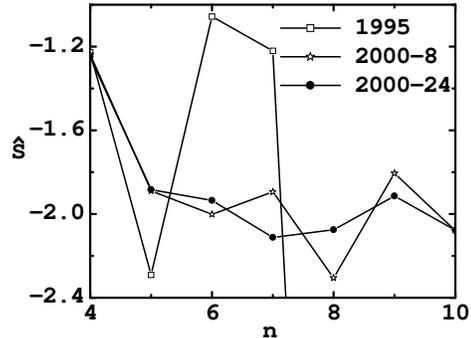}
\caption{$\widehat{S}_m$ versus $m$.}
\label{fig:exslo}
\end{figure}

The corrections of order $(1/m)^2$ in $r_m$ introduce an uncertainty of order $1/m$ in the estimation of 
$\alpha$.
It is possible to eliminate this effect by 
using \cite{nickel80}
the  
extrapolated slope ($\widehat{S}_m$). For this purpose, we first introduce 
the normalized slope $S_m$ defined a 
\begin{equation}
S_m  = -m(m-1)(r_m - r_{m-1})/(mr_m -(m-1)r_{m-1})
\end{equation}
The values of $S_m$ are shown in Fig. \ref{fig:noslo} again showing a much better stability for the 
2000 series. The volume effects are smaller than for $\widehat{R}_m$. They are typically of order 0.01 or smaller. It is known \cite{nickel80} that $S_m\simeq \alpha-2 +K/m$. Using a fit of this form, 
we obtain $\alpha=-0.03$ for the $m=4$ to 10 values.
One can then remove the $1/m$ corrections by using \cite{nickel80} the extrapolated 
slope $\widehat{S}_m$ defined as 
\begin{equation}
\widehat{S}_m  =  mS_m-(m-1)S_{m-1}\ .
\label{eq:shat}\end{equation}
It provides an estimator with corrections of the form
\begin{equation}
\widehat{S}_m =\alpha -2  -B m^{-\Delta }+O(m^{-2})\ ,
\label{eq:convdecay} \end{equation}
where $\Delta$ is an exponent corresponding to possible nonalytical corrections to the leading term 
Eq. \ref{eq:convpar}. In 3 dimensional scalar models $\Delta$ is related to irrelevant directions.
The values of $\widehat{S}_m$ are shown in Fig. \ref{fig:exslo}. Again, the fluctuations 
are larger than for the unextrapolated quantity and the 2000 data is much more stable. 
For the 2000 data on a $24^4$ lattice, we have $\widehat{S}_9=-1.91$ $\widehat{S}_{10}\simeq -2.08$. Based on other estimates, we conclude that $|\alpha|<0.1$.

In summary, the analysis of ratios suggests that $P$ has a non-analytic, power-like singularity 
\begin{equation}
P_{n.-a.}\propto (1/5.7(1)-1/\beta)^{1.0(1)}\ .
\label{eq:pna}
\end{equation}

\section{Remarks}
\label{sec:remarks}

It is not difficult to find an asymptotic series (which has a zero radius of convergence) for which the beginning coefficients suggest a power singularity. A simple example is 
\begin{equation}
Q(\beta)=\int_0^{\infty}dt {\rm e}^{-t}t^{B}[1-t\beta_c/(B \beta)]^{1-\alpha}	
\end{equation}
the ratios of coefficients of $\beta^{-1}$ can be calculated exactly: 
\[
r_m=((B +m)/B)\beta_c(1+(\alpha -2 )/m).
\] 
For $B $ sufficiently large and $m<<B$, we have 
\[r_m\simeq \beta_c(1+(\alpha -2 )/m), \] which corresponds approximately to the situation observed 
for the series discussed above. On the other hand, 
for $m>>B$ we have $r_m \propto m$ whichs shows 
that for sufficiently large $m$, the coefficients grow factorially.

It is also well-known that the convergence of perturbative series can be improved by using the tadpole improvement \cite{lepage92}.
One defines a new series
\begin{equation}
P \simeq \sum_{m=0}^{K} b_m \beta^{-m}=\sum_{m=0}^{K} e_m \beta_R^{-m} + O(\beta_R^{-K-1})
\end{equation}
with a new expansion parameter
\begin{equation}
\beta_R^{-1}=\beta^{-1}\frac{1}{1-\sum_{m=0} b_m \beta^{-m}} \ .
\end{equation}

\begin{table}[t]
\begin{tabular}{||c|c|c||}
\hline
 $m$&$b_m$&$e_m$\cr
 \hline
1 & 2 & 2 \cr 2 & 1.2208 & 
    -2.779 \cr 3 & 2.9621 & 3.637 \cr 4 & 9.417 & 
    -3.961 \cr 5 & 34.39 & 4.766 \cr 6 & 136.8 & 
    -3.881 \cr 7 & 577.4 & 6.822 \cr 8 & 2545 & 
    -1.771 \cr 9 & 11590 & 17.50 \cr 10 & 54160 & 
   48.08 \cr 
\hline
\end{tabular}
\caption{$b_m$: regular coefficients;  $e_m$: tadpole improved coefficients}
\end{table}
The new coefficients are shown in Table I. If we exclude the last coefficient which may not be very reliable, we find that series is alternate and has  much smaller coefficients.
The first six ratios are of order -1, which suggests a singularity at negative $\beta_R$. 
A discontinuity in $P$ at $\beta\simeq -22$ was observed in \cite{gluodyn04}, however 
we have no interpretation for $\beta_R<0$.

\section{Direct Search for a singularity}
\label{sec:direct}
If we take Eq. (\ref{eq:pna}) at face value, it implies that the third derivative of the 
free energy density should have a singularity with an exponent close to -1. More precisely, we would expect:
\begin{equation}
\partial^2P / \partial \beta^2 \propto (1/5.7(1)-1/\beta)^{-1.0(1)}\ .
\label{eq:ddpna}
\end{equation}
For symmetric finite lattices with $L^D$ sites and periodic boundary conditions, the number of plaquettes is 
\begin{equation}
\mathcal{N}_p\equiv\ L^D D(D-1)/2\ .
\end{equation}
Using the free energy density
\begin{equation}
f\equiv-(1/\mathcal{N}_p)\ln Z\ ,
\end{equation}
we define the average plaquette
\begin{equation}
	P=\partial f/\partial \beta =(1/\mathcal{N}_p)\left\langle \Sigma\right\rangle
\end{equation}
with
\begin{equation}
\Sigma\equiv \sum_{p}(1-(1/N)Re Tr(U_p)) \ .
\end{equation} 
We also define the higher moments:
\begin{equation}
-	\partial P / \partial \beta =(1/\mathcal{N}_p)[\left\langle \Sigma ^2\right\rangle- \left\langle \Sigma\right\rangle^2]\ ,
\end{equation}
and 
\begin{equation}
	\partial^2 P /\partial \beta^2 =(1/\mathcal{N}_p)[\left\langle \Sigma ^3\right\rangle-3 \left\langle \Sigma\right\rangle \left\langle \Sigma^2\right\rangle+ \left\langle \Sigma\right\rangle^3]
\end{equation}

It should be noted that there is some loss of precision in the calculation of the 
higher moments. For instance, in $-	\partial P\/\partial \beta$, the two terms are 
of order $\mathcal{N}_p$ but their difference is of order 1. 
For $\partial^2 P/\partial \beta^2$, the three terms are of order $\mathcal{N}_p^2$ while 
their combination is of order 1. For a symmetric lattice with $10^4$ sites, for instance, the second derivative of the plaquette will appear in the ninth significant digit and the use of double precision is crucial. To give an idea, 400,000 reasonably uncorrelated configurations 
were necessary in order to get a signal significantly larger than the statistical fluctuations 
for the four points on a $10^4$ lattice in Fig. \ref{fig:pd2}. For $\beta > 5.7$, the second derivative seems to be of the same order or smaller than the fluctuations for the same volume.

\begin{figure}
\includegraphics[width=2.8in,angle=0]{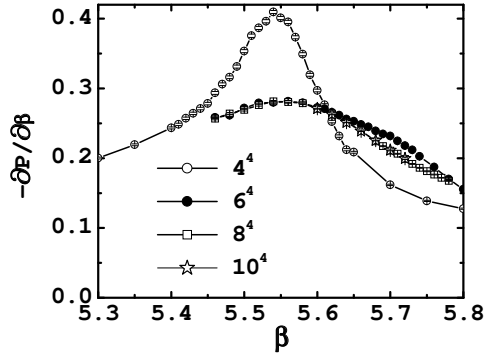}
\caption{First derivative of $P$ versus $\beta$.}
\label{fig:pd1}
\end{figure}

\begin{figure}
\includegraphics[width=2.8in,angle=0]{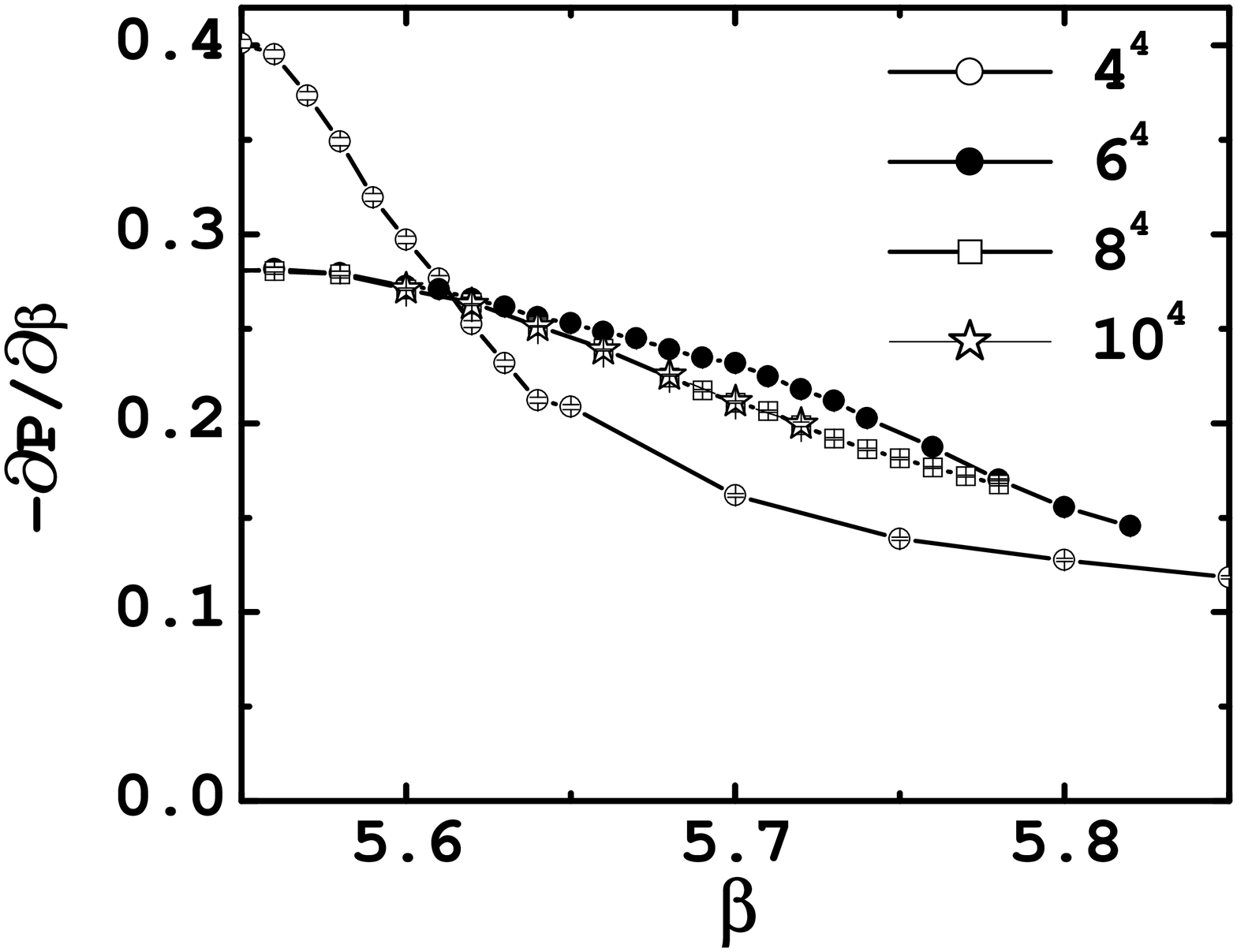}
\vskip-5pt
\includegraphics[width=2.8in,angle=0]{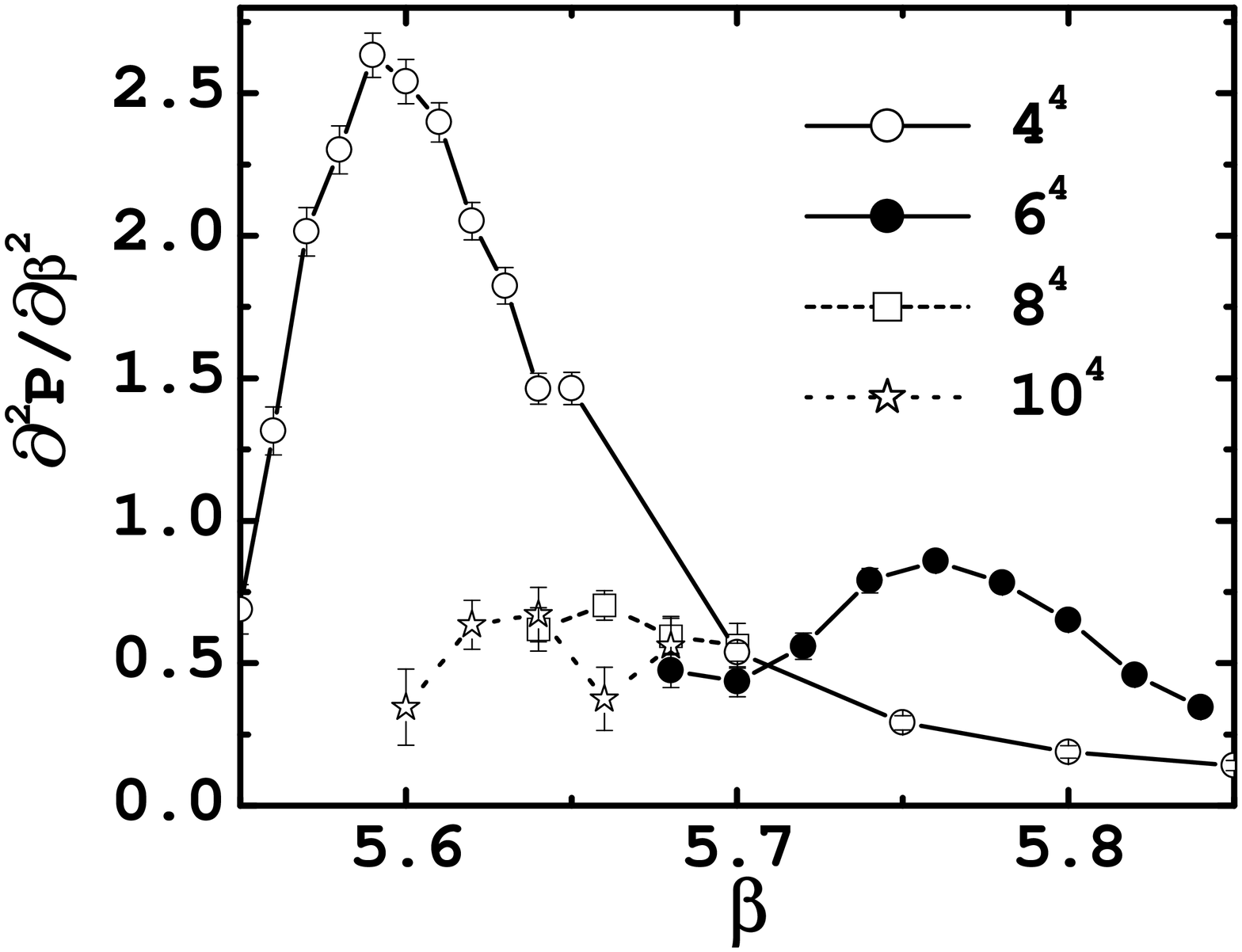}
\caption{First and second derivative of $P$ versus $\beta$.}
\label{fig:pd2}
\end{figure}

We have calculated the plaquette and its first two derivatives on $L^4$ lattices, with $L$= 4, 6, 8 and 10. 
The first derivative is shown in Fig. \ref{fig:pd1}. When going from $L=4$ to $L=6$, the peak moves right and its height diminishes. This situation is very close to Fig. 1 
of Ref. \cite{engels96} in the case of $SU(2)$. When increasing $L$ to 8, the values around 5.7 slighly drops and then stabilizes for $L=10$. 

The second derivative of the plaquette  is shown in Fig. \ref{fig:pd2}. For comparison, we have also reproduced the first derivative over the same range and with the same scale. In general, we found a reasonable agreement between the values obtained by subtracted averages or by taking the 
a numerical derivative as found in \cite{engels96}. The figure makes clear that peak seems to disapear to a level close to our statistical fluctuations when the volume increases.

\begin{figure}
\includegraphics[width=3.8in,angle=0]{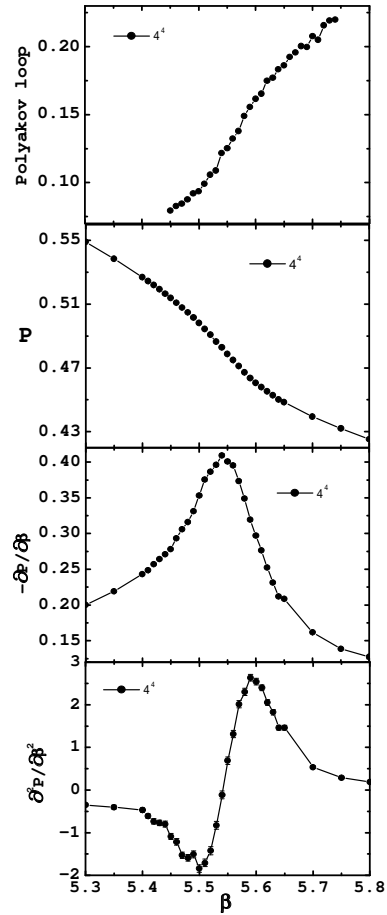}
\caption{The Polyakov loop, $P$, first and second derivative of $P$ versus $\beta$ for 
a $4^4$ lattice.}
\label{fig:4to4}
\end{figure}

\begin{figure}
\includegraphics[width=3.8in,angle=0]{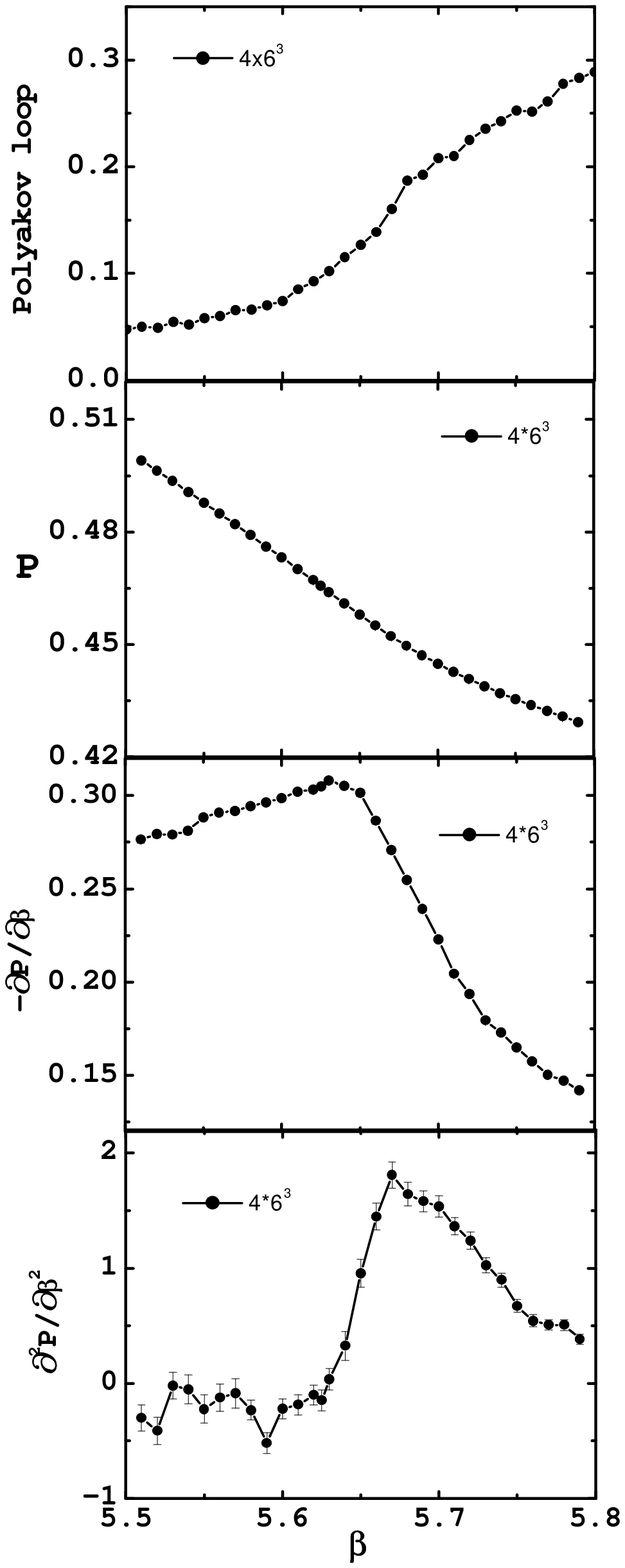}
\caption{The Polyakov loop, $P$, first and second derivative of $P$ versus $\beta$ for a $4\times 6^3$ lattice.}
\label{fig:46to3}
\end{figure}

\begin{figure}
\includegraphics[width=3.8in,angle=0]{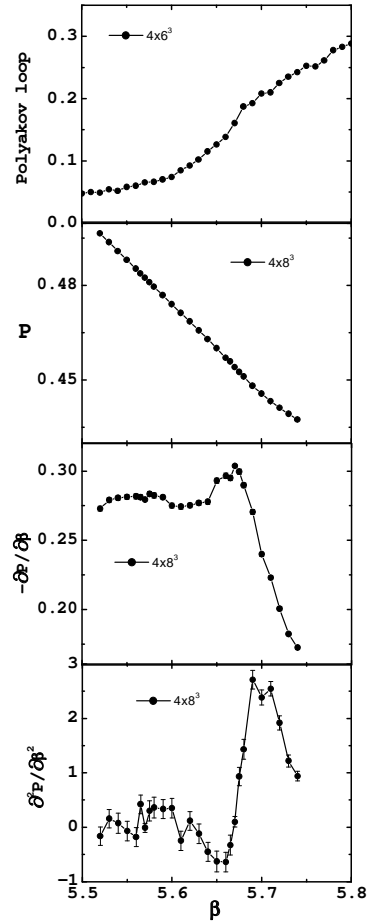}
\caption{The Polyakov loop, $P$, first and second derivative of $P$ versus $\beta$ for 
a $4\times 8^3$ lattice .}
\label{fig:48to3}
\end{figure}

We have also calculated the same quantities on $4L^3$ lattices with $L=$ 4,6 and 8.
The results are shown in Figs. \ref{fig:4to4}, \ref{fig:46to3} and \ref{fig:48to3}.
The sudden jump in the second derivative of $P$ persists when $L$ increases.
The location of the jump coincides with the onset of the average of the Polyakov loop. The critical values of $\beta$ corresponding to a first order finite 
temperature transition are well-known for $4L^3$ lattices \cite{alves91,iwasaki92,boyd96}. For $L$ sufficiently large, it occurs very close 
to $\beta=5.69$. The locations of the jumps seen in the second derivative in Figs. 
7-9 follow very closely the quantity $\beta_x^0$, the real part of the leading 
zero of the partition function, of Table 9 of Ref. \cite{alves91} (5.552 for $L=4$, 5.650 for $L=6$ and 5.674 for $L=8$).

The fact that for a 
$4L^3$ lattice, $\beta_c\simeq 5.69$ is very close to $\beta_c\simeq 5.74$ obtained with the extrapolated ratio from the perturbative series seems to be 
a coincidence. Had we taken $N_t$ to be 12, we would have obtained \cite{boyd96} $\beta_c\simeq 6.33$.
Note that on $L^4$ lattices, the onset of the Polyakov loop occurs at larger values of $\beta$. For instance, on a $8^4$ lattice, it is 0.1 near $\beta = 6.6$. However, we found nothing but a smooth decrease for the first derivative of $P$ in this range.

\section{Resolution of the paradox}
\label{sec:resolution} 

It is possible to regularize the singularity that, according to Eq. (\ref{eq:convpar}), would appear in $\partial^2 P/\partial \beta^2$ by replacing the singularity at $1/\beta_c$ by a pair of complex conjugated singularities in the $1/\beta$ complex plane. If we denote the imaginary 
part of the location of the singularity by $\Gamma$, then some apparently difficult compromise needs to be achieved. On one hand, if $\Gamma$ is too large, it produces periodic modulations of the ratios. Fig. \ref{fig:rat} gives no hint of such an effect. On the other hand, if $\Gamma$ is too small it does not fulfill its purpose. The question is thus if we can find a window of acceptable values for $\Gamma$.

A simple alternative to Eq. (\ref{eq:convpar}) can be designed by assuming that the critical 
point in the fundamental-adjoint plane has mean field exponents \cite{heller95} and in particular $\alpha=0$. We will further 
assume an approximate logarithmic behavior 
\begin{equation}
-\partial P/\partial \beta \propto {\rm ln}((1/\beta_m-1/\beta)^2+\Gamma^2)\ ,
\end{equation}
on the axis where the adjoint term of the 
action is zero (the range of parameters considered here). 
$1/\beta_m$ denotes the value where the argument of the logarithm is maximal on this axis.
This implies the approximate form 
\begin{equation}
\partial^2 P/\partial \beta^2\simeq - C \frac{(1/\beta_m-1/\beta)}{\beta^3 ((1/\beta_m-1/\beta)^2+\Gamma^2)}
\label{eq:resol}
\end{equation}
The $\beta^3$ at the denominator ensures that the series starts at $\beta^{-3}$.
The three unknown parameters can be approximately fixed by minimizing some linear combination 
(with positive ``weights'') 
of the square of the differences
between the coefficients of the series Eq. (\ref{eq:resol}) and the actual ones. 
We expect the coefficients of larger order to carry more information about the non-analytic behavior, consequently, more weight should be given to the large order coefficients.  
For instance 
if we minimize a $\chi^2$ which is the sum of the square of the relative errors for the last four coefficients, we obtain $\beta_m\simeq 5.78$, 
$\Gamma \simeq 0.0058$, and $C\simeq 0.15$. Integrating twice, we obtain the approximate series 
with coefficients which can be compared with the $c_m$ given in Table II:
\begin{eqnarray}
 P\simeq 0.44\beta^{-1}&+&  
    0.85 \beta^{-2}+ 2.44  \beta^{-3}+
    8.42 \beta^{-4}\cr 
    +32.27 \beta^{-5}
    &+&
    132.3 \beta^{-6}+568.9\beta^{-7}\cr +
    2533 \beta^{-8}&+&11590\beta^{-9}+
   54160\beta^{-10}\ .
\end{eqnarray}   
The agreement increases with the order. 
The relative error on the fifth coefficient is 
about 6 percent and keep decreasing to a level smaller than the numerical errors as the order increases. We have tried with many other weights involving the last 6 coefficients and found very little variations in the estimation of $C$ (typically $|\delta C|\sim 0.01$) and $\beta_m$ (typically $|\delta \beta_m|\sim 0.02$). On the other hand, $\Gamma$ varies more rapidly under changes of the weights in the $\chi^2$ function. We found values of $\Gamma$ between 0.003 and 0.007.

The stability of $C$ and $\beta_m$ can be used to set a lower bound on $\Gamma$. Given that the approximate form of $\partial ^2 P/\partial \beta^2$ in Eq. (\ref{eq:resol}) has extrema at $1/\beta =1/\beta_m \pm \Gamma$. As we do not observe values larger than 0.3 near $\beta =5.75$ (see Fig. \ref{fig:pd2}) we get the approximate bound 
\begin{equation}
\frac{C}{2\beta_m^3\Gamma}<0.3
\end{equation}
This implies the lower bound $\Gamma>0.001$. 
On the other hand, large values of $\Gamma$ are also excluded. As we never found estimate close to 0.01, we conclude that 
\begin{equation}
0.001<\Gamma<0.01 \ .
\end{equation}

We also performed calculations with an assumption similar to Eq. (\ref{eq:resol}) but with  $((1/\beta_m-1/\beta)^2+\Gamma^2)^{1+(\alpha/2)}$ at the denominator, 
for small positive and negative values of $\alpha$. We found very similar ranges of values 
for the unknown parameters and we were able to draw very similar conclusions as for $\alpha =0$. 

A puzzling aspect of Fig. \ref{fig:pd2} is that the maximum of the first and second derivatives are not located near 5.78 but near lower values (5.55 and 5.63 repectively). 
Using the parameters obtained with the procedure described above, 
we found that a typical value for the maximum of the second derivative of $P$ is 0.1 or below. 
This is clearly below our numerical resolution and below the maximum value 0.6 in Fig. \ref{fig:pd2}. This can be explained from the fact 
that the perturbative series becomes a poor estimate of P when 
$\beta$ becomes too small. 
Using  the parametrization of this difference (the ''non-perturbative part of the plaquette'') of Ref.  \cite{rakow2002} and taking two derivatives, we estimate that the nonperturbative contribution to  $\partial^2  P/\partial \beta^2$ can be approximately written as $2.84\times 10^{11}\times {\rm exp}({-\frac{16\pi^2\beta}{33}})$ for $5.6<\beta<5.8$. This function takes the values 
0.25 at $\beta=5.8$ and dominates the perturbative part. It also takes the value 0.65 at $\beta=5.6$ which is consistent with our numerical calculation. As the parametric form of the 
nonperturbative contribution is still being debated, this is not the last word on the question, however, it seems clear that the nonperturbative part plays a major role in explaining Fig. 
\ref{fig:pd2}.

\section{Consistency checks}
\label{sec:checks}

It would be interesting to check if the complex singularities suggested by the present analysis could be seen using independent methods. In this section, we discuss the possibility of 
using the strong coupling expansion and the zeroes of the partition function for this purpose.
The strong coupling expansion of the free energy has been calculated up to order 16 in Ref. \cite{balian74err} for $SU(2)$ and $SU(3)$. With a suitable rescaling of $\beta$, these 
series can be used to calculate $P$ and its derivatives in power of $\beta$ (with the 
normalization of Eq. (\ref{eq:z})).
In the case of $SU(2)$, the complex singularities of the specific heat were discussed in various sets of complex variables that have a simple interpretation 
for small values of $\beta$ \cite{falcioni80,falcioni81} and compared with zeroes of the partition function on a $4^4$ lattice \cite{falcioni82}. These zeroes were calculated using 
a re-weighting of Monte Carlo data at real $\beta$.

In order to provide a comparison, we will start with the case of $SU(2)$ which is better understood. 
We constructed the expansion in powers of $\beta$ of $P$ using Ref. \cite{balian74err}.
This serie is 1 plus an odd series in $\beta$. This can be seen as a consequence of the 
identity $P(\beta)+P(-\beta)=2$ derived in Ref. \cite{gluodyn04}. The sign of the nonzero coefficients alternates. The ratio of successive odd coefficients seems to converge toward a value close to 
$-0.4$. 
This indicates a singularity at $\beta^2\simeq -2.5$, or in other words, a pair of purely imaginary singularities at $\beta \simeq \pm i 1.6$. We found 36 Pad\'e approximants of $P$ with their singularity closest to the origin located nearby (within 0.1 in both coordinates) this estimate. 
More sophisticated estimators such as the ones of Sec. \ref{sec:seran} do not improve the accuracy of the estimation.
In \cite{falcioni80}, the single loop variable $J$ was used instead of $\beta$. Using a ratio analysis of the series in $J$, these authors concluded that there is a singularity near 
$J^2\simeq -1/6$. This corresponds to $\beta \simeq \pm i 1.5$, in good agreement 
with the estimate we gave above. 

We have inspected the poles of the Pad\'e approximants for $P$ and its first two derivatives in a horizontal strip defined by the conditions $Re \beta >1.5 $ and $|Im \beta|<1$. We found single poles on the real axis and pairs of conjugated poles rather far from the real axis. To fix the ideas, for the second derivative of $P$, the pair closest to  the real axis in this strip  is $1.79\pm i0.43$ for a $[3/9]$ approximant. Also, the poles on the real axis tend to cluster. For instance, for the first derivative of $P$ there are 24 approximants with a pole between 2.1 and 2.5, and for the second derivative, 13 approximants with a pole between 1.7 and 1.8.

In Ref. \cite{falcioni80}, an attempt was made to find evidence for complex singularities in a new  variable $z\equiv 6J^2/(6J^2+1)$. It was concluded that the existing series was too short to see a clear departure from a singularity on the real axis (near $\beta =2.2$). On the other hand, complex zeroes of the 
partition function were found \cite{falcioni82} on $4^4$ lattice near $\beta =2.2+\pm 
i 0.15$. It is plausible that the imaginary part increases with the volume 
and possibly reaches the estimate   
$\beta \simeq 2.2+\pm 
i 0.3$  based on a more sophisticated treatment of the high-temperature expansion \cite{falcioni81}. We are not aware of any numerical check of this statement on larger lattices.
However, the fact that the two independent estimates are close and that one expects the specific heat peak to broaden with the volume, makes the existence of a singularity at $\beta = 2.2(1) \pm i 0.2(1)$ quite plausible. Consequently, it seems that the Pad\'e approximants often provide a good estimation 
for the location of the singularities close to the origin but not for the expected singularities close to the real axis that are farther from the origin.

For $SU(3)$, we analyzed the $\beta$ expansion of $P$ following a similar procedure.
With the exception of the third coefficient, the coefficients are nonzero. The signs do not 
show any obvious periodic pattern. 
If we plot the logarithm of the absolute value of the coefficients, we see that they approximately fall on a line, however there are exceptions. We already mentioned the vanishing third coefficient, in addition, the 11th (15th) coefficient 
falls significantly below (above) the linear fit. The slope of the linear fit is -1.49 if the 
coefficients 4 to 15 are used. This indicates a radius of convergence of approximately 4.5. Due to the significant discrepancies from the linear behavior, the ratio analysis does not reveal any obvious information. 

The poles of the Pad\'e approximants are quite irregular. However, many approximants have a pair of complex conjugated poles near $5{\rm e}^{\pm i\pi/3}$. 
We have found 12 Pad\'e approximants of $P$ for which the poles closest to the origin were located close to these two points (within 0.3 in each coordinate).
An inspection of the poles of the Pad\'e approximants for $P$ and its first two derivatives in a horizontal strip defined by the conditions $Re \beta >4$ and $|Im \beta|<2$ shows single poles on the real axis and pairs of conjugated poles far from the real axis. 
For instance, for the second derivative of $P$, the pair closest to  the real axis in this strip is $4.74\pm i 1.72$ for a $[5/5]$ approximant. Considering that in the case of $SU(2)$ we were 
unable to find the expected singularity near $\beta \simeq 2.2+\pm 
i 0.2$, it seems plausible that a similar phenomenon occurs for $SU(3)$. It would be interesting to repeat the above analysis using changes of variables as in the $SU(2)$ case.

In Sec. \ref{sec:direct}, we have already mentioned the fact that the leading zeroes of the partition function for $SU(3)$ on a $4L^3$ lattice had been calculated \cite{alves91}. On a $4^4$ lattice, they are located near $\beta=5.55\pm i 0.12$. On larger symmetric $L^4$ lattices, we expect that the imaginary part will increase with $L$. We are not aware of numerical calculations for $L>4$. 
With the parametrization of Sec. 
\ref{sec:resolution}, we predict complex singularities at $\beta=\beta_m\pm i \beta_m^2\Gamma$. 
For $\beta_m=5.75$ and $0.001<\Gamma<0.01$, we predict an imaginary part between $0.03$ and $0.3$. It seems feasible to check this prediction for $8^4$ or larger lattices, using the method documented in Ref. \cite{alves91}. 

It should also be noted that new methods have been developed to determine the order of a phase transition using the distribution of zeroes of the partition function and their impact angle in the reduced variables complex plane \cite{janke05}. It would be quite interesting to apply 
these methods to interpret the variations of zero distribution when an adjoint term is added to the action. 

\section{Conclusions}
We have shown that the apparent conflict between an hypothetical singularity in the second 
derivative of $P$ suggested by the perturbative series  and the absence of  evidence for a peak with height increasing with the volume  
on isotropic lattices, can be resolved by moving the singularity in the complex $1/\beta$ plane. If the imaginary part of the location of the singularity $\Gamma$ is within the range $0.001< \Gamma < 0.01$, it is possible to limit the second derivative of $P$ within an acceptable range without affecting drastically the behavior of the perturbative coefficients.  This picture seems consistent with a small value of $\alpha$ but our numerical analysis does not single out the 
mean field value $\alpha =0$. 

It would be interesting to calculate the zeroes of the partition function for $SU(3)$ on $8^4$ or larger 
symmetric lattices using the the re-weighting \cite{falcioni82,alves91} method. We expect that as the volume increases, the locations of the leading zeroes should stabilize at values  $5.7(1)\pm i 0.2(1)$. Changes of variables analog to those used in $SU(2)$ \cite{falcioni80,falcioni81} should be used in order to see if there is agreement with the other methods.
It would also be interesting to follow the distribution of zeroes and their impact angle when an adjoint term is added using the results of Ref. \cite{janke05}.

\begin{acknowledgments}
This work was completed while Y. M. was visiting the Kavli Institute for Theoretical 
Physics. Y. M. thanks the organizers of the workshop ''Modern Challenges for Lattice Field Theory'' for making possible many conversations about old and new topics related 
to the questions discussed here. 
Y. M. aknowledges valuable discussions with the participants. 
We thank G. Parisi and M. Creutz for comments regarding third order phase transitions. 
We used the code FermiQCD \cite{dipierro05}  for our numerical calculations.
This 
research was supported in part by the National Science Foundation
under Grant No. PHY99-07949, 
in part by the Department of Energy
under Contract No. FG02-91ER40664. 
\end{acknowledgments}

\end{document}